\documentclass[twocolumn,preprintnumbers,prl,superscriptaddress]{revtex4-1}

\usepackage{graphicx}
\usepackage{dcolumn}
\usepackage{bm}

\usepackage[colorlinks=true,linkcolor=black, citecolor=black,
urlcolor=black]{hyperref}

\usepackage{relsize}
 \usepackage{multirow,graphics}
 \usepackage{amstext}
 \usepackage{amssymb}
 \usepackage{amsmath}
 \usepackage{graphicx}
 \usepackage{color}
 \usepackage{dsfont} 

 \usepackage[caption=false]{subfig}

\newcommand{\p}{\partial}
\newcommand{\Tr}{{\text{Tr}}}

\DeclareMathOperator{\idop}{\mathds{1}}
\newcommand{\sfrac}[2]{{\textstyle\frac{#1}{#2}}}
\newcommand{\half}{\sfrac{1}{2}}
\newcommand{\ihalf}{\sfrac{i}{2}}

\newcommand{\gen}[1]{\mathrm{#1}}
\newcommand{\comm}[2]{[#1,#2]}

\newcommand{\dd}{\mathrm{d}}

\newcommand{\alg}[1]{\mathfrak{#1}}
\newcommand{\mon}{\mathrm{T}}
\newcommand{\lax}{\mathrm{L}}
\newcommand{\permop}{\mathbb{P}}
\newcommand{\rmat}{\mathrm{R}}
\newcommand{\abs}[1]{|#1|}

\makeatletter
\newlength{\apb@width}
\newcommand{\autoparbox}[2][c]{\settowidth{\apb@width}{#2}\parbox[#1]{\apb@width}{#2}}
\newcommand{\includegraphicsbox}[2][]{\autoparbox{\includegraphics[#1]{#2}}}
\makeatother

\makeatletter
\def\mr@ignsp#1 {\ifx\:#1\@empty\else #1\expandafter\mr@ignsp\fi}%
\newcommand{\multiref}[1]{\begingroup
\xdef\mr@no@sparg{\expandafter\mr@ignsp#1 \: }%
\def\mr@comma{}%
\@for\mr@refs:=\mr@no@sparg\do{\mr@comma\def\mr@comma{,}\ref{\mr@refs}}%
\endgroup}
\makeatother

\newcommand{\hypref}[2]{\ifx\href\asklfhas #2\else\href{#1}{#2}\fi}

\newcommand{\tabref}[1]{Tab.~\multiref{#1}}

\newcommand{\figref}[1]{Fig.~\multiref{#1}}
\renewcommand{\eqref}[1]{(\multiref{#1})}

\newcommand{\rvec}[1]{\reflectbox{\ensuremath{\vec{\reflectbox{\ensuremath{#1}}}}}}

\begin{document}

\preprint{MITP/17-049, HU-EP-17/20}

\title{Yangian Symmetry for Fishnet Feynman Graphs}

\author{Dmitry  Chicherin}  
\email{chicherin@uni-mainz.de}
\affiliation{%
PRISMA Cluster of Excellence, Johannes Gutenberg University, 55099 Mainz, Germany
}%
\author{Vladimir Kazakov} 
\email{kazakov@physique.ens.fr} 
\affiliation{%
Laboratoire de Physique Th\'eorique, D\'epartement de Physique de l'ENS, Ecole Normale Sup\'erieure, PSL Research University, Sorbonne Universit\'es,  UPMC Univ. Paris 06, CNRS, 75005 Paris, France
}%
\author{Florian Loebbert}
\email{loebbert@physik.hu-berlin.de} 
\affiliation{%
Institut f\"ur Physik, Humboldt-Universi\"at zu Berlin,
Zum Gro{\ss}en Windkanal 6, 12489 Berlin, Germany
\&\\
Kavli Institute for Theoretical Physics,
University of California, 
Santa Barbara, CA 93106, USA
}%
\author{Dennis M\"uller} 
\email{dmueller@physik.hu-berlin.de}
\affiliation{%
Institut f\"ur Physik, Humboldt-Universi\"at zu Berlin,
Zum Gro{\ss}en Windkanal 6, 12489 Berlin, Germany
\&\\
Kavli Institute for Theoretical Physics,
University of California, 
Santa Barbara, CA 93106, USA
}%
\author{De-liang Zhong}%
 \email{zdlzdlzdl@gmail.com}
\affiliation{%
Laboratoire de Physique Th\'eorique, D\'epartement de Physique de l'ENS, Ecole Normale Sup\'erieure, PSL Research University, Sorbonne Universit\'es,  UPMC Univ. Paris 06, CNRS, 75005 Paris, France
}%


\date{\today}

\begin{abstract}
Various classes of fishnet Feynman graphs are shown to feature a Yangian symmetry over the conformal algebra. We explicitly discuss scalar graphs in three, four and six spacetime dimensions as well as the inclusion of fermions in four dimensions.
The Yangian symmetry results in novel differential equations for these families of largely unsolved Feynman integrals. 
Notably, the considered fishnet graphs in three and four dimensions dominate the  correlation functions and scattering amplitudes in specific double scaling limits of planar, $\gamma$-twisted   $\mathcal{N}=4$ super Yang--Mills or ABJM theory. Consequently, the study of fishnet graphs allows us to get deep insights into  the integrability of the planar AdS/CFT correspondence.
\end{abstract}

                              
\maketitle


\section{Introduction}\label{sec:intro}

Feynman diagrams represent the main tool for the  study of complex physical phenomena---from fundamental interactions of elementary particles to diverse solid state systems.  
In spite of the great progress in computing individual Feynman graphs with multiple loop integrations, 
examples of exact all-loop results for important physical quantities (such as amplitudes, correlators, etc.)   are rare in dimensions greater than two. 
Remarkably, there exist certain types of  planar graphs
with a particularly regular structure, which may be calculable at any loop order. Examples are the regular tilings of the two-dimensional plane. These diagrams become accessible due to their integrability properties, in close analogy to the quantum integrable one-dimensional Heisenberg spin chains.  Apart from providing new, powerful methods for the computation of large classes of particular Feynman graphs, these observations reveal the  interplay between various physical systems and a rich variety of mathematical aspects related to quantum integrability.                                   

\begin{figure}
\includegraphics[scale=.7]{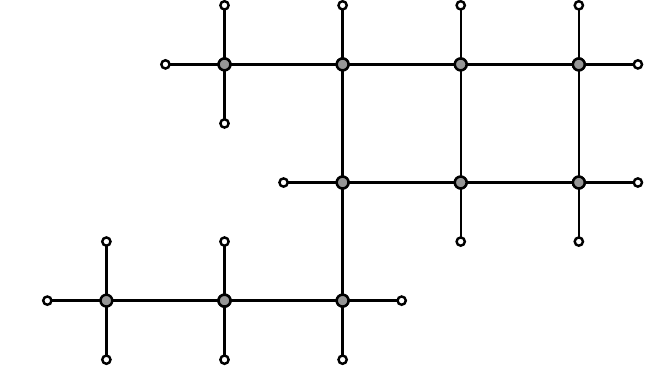}
\caption{
Example of a conformal scalar fishnet Feynman graph in four dimensions. Filled blobs denote loop integrations, white blobs represent external points $x_k$.}
\label{fig:fishnet}
\end{figure}

A prime example in the above class of Feynman graphs are scalar fishnets in four dimensions, built from four-point vertices connected by massless propagators (cf.\ \figref{fig:fishnet}). These represent one of the three regular tilings of the Euclidean plane and,
except for the simplest example, solving this class of Feynman integrals for generic external parameters is an open problem. On the other hand, these square fishnet graphs are subject to outstanding properties: 
Firstly, they feature a (dual) conformal Lie algebra symmetry, which makes it natural to express them using conformal cross ratios.
They are finite, i.e.\ free of IR or UV divergencies, such that their conformal symmetry is unbroken for generic external kinematics.
Moreover, already in 1980 A.~Zamolodchikov demonstrated that scalar fishnet graphs can be interpreted as integrable vertex models~\cite{Zamolodchikov:1980mb}.
Furthermore, in the planar limit fishnet graphs appear to dominate physical quantities, such as  scattering amplitudes and correlators,  of the bi-scalar CFT recently found by \"O.~G\"urdogan and one of the authors~\cite{Gurdogan:2015csr}
as  a specific double-scaling limit of the integrable  $\gamma$-twisted $\mathcal{N}=4$ SYM theory. This non-unitary bi-scalar CFT is defined by the Lagrangian
 \begin{align}
    \label{eq:Lagr}
    {\cal L}_\phi= N_\text{c}\Tr
    \big(\p^\mu\phi^\dagger_1 \p_\mu\phi_1+\p^\mu\phi^\dagger_2 \p_\mu\phi_2+\xi^2\,\phi_1^\dagger \phi_2^\dagger \phi_1\phi_2\big).
  \end{align}
Its basic physical quantities (anomalous dimensions, correlation functions etc.) are determined by a very limited number of Feynman graphs at each loop order and efficiently calculable via integrability \cite{Caetano:2016ydc,Gromov:2017cja}.   

In this letter we add a further remarkable property to the above list of features of fishnet graphs. We demonstrate that their conformal symmetry extends to a non-local Yangian symmetry. 
 This symmetry yields novel differential constraint equations for this class of Feynman integrals. 
 
A single scalar fishnet graph of the above type represents a single-trace correlator
of the bi-scalar theory \begin{equation}\label{eq:tr-corr}
K(x_1,\dots,x_{n})=\langle\mathrm{Tr}[\chi_1(x_1)\dots \chi_n(x_n)]\rangle.
\end{equation}
Here $\chi_k\in \{\phi_1,\phi_2,\phi_1^\dagger,\phi_2^\dagger\}$ and $x_i$ is the spacetime coordinate of the field $\chi_i$.
Importantly, via the relation to this bi-scalar model, we define a CFT  for the all-loop study of Yangian-invariant correlators and scattering amplitudes, similar to those appearing in the AdS/CFT duality. 
After having discussed the above scalar fishnets in four dimensions, we will show that the class of Yangian-invariant Feynman graphs is actually much richer and extends to different dimensions, particle species and more exotic tilings of the plane.


\section{The Box and the Yangian}

\begin{figure}
 \captionsetup[subfloat]{farskip=0pt,captionskip=0pt}
 \subfloat[\label{fig:boxlhs}]{
 \includegraphics[scale=1]{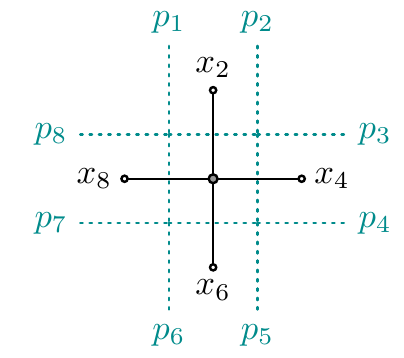}
}
\subfloat[\label{fig:boxrhs}]{
\includegraphics[scale=1]{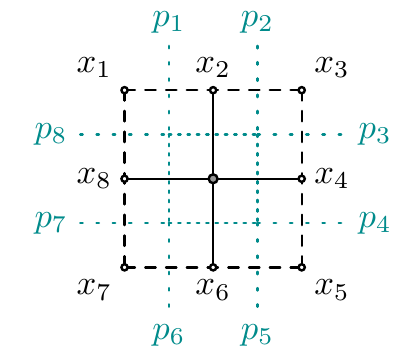}
}
\caption{The box integral in momentum (black) and dual (green) coordinate space. It will be convenient to distinguish (a) off-shell  and (b) on-shell  external momenta.
Note the relabelling of coordinates with respect to \protect\eqref{eq:boxint}.}
\label{fig:box}
\end{figure}

The most elementary representative in the class of fishnet graphs is the scalar box integral
\footnote{Here we do not discuss the even simpler tree-level diagrams. See \cite{Chicherin:2017cns} for comments.},
 cf.\ \figref{fig:box}. In fact, this integral is the only member of this family, which has been solved explicitly. It is conveniently written in terms of variables $x_i$ which can be related to dual momenta via $p_i^\mu=x_i^\mu-x_{i+1}^\mu$. The scalar box integral then reads
 \begin{equation}\label{eq:boxint}
 I_4=\int \dd^4 x_0 \frac{1}{x_{01}^2x_{02}^2x_{03}^2x_{04}^2},
 \end{equation}
 and evaluates to a combination of logs and dilogs of conformal cross ratios \cite{ussyukina1993approach}.
 The above box integral \eqref{eq:boxint}---as well as all fishnet graphs composed from such elementary boxes---are invariant under the conformal algebra $\alg{so}(2,4)$. On a generic scalar fishnet graph, the conformal generators are represented via their usual tensor product representation $\gen{J}^A =\sum_{j=1}^n\gen{J}_j^A$ with the index \(j\) labeling the external legs of the graph and  the index $A$ enumerating the following differential operators:
 \begin{align}\label{eq:lev0gens}
 \gen{D}&=-i x_\mu \partial^\mu-i\Delta,
 &
\gen{ L}_{\mu\nu}&=ix_\mu \partial_\nu-ix_\nu \partial_\mu,
\\
 \gen{P}_\mu &=-i\partial_\mu,
 &
\gen{ K}_\mu&= i x^2 \partial_\mu -2 ix_\mu x^\nu \partial_\nu -2 i \Delta x_\mu.\nonumber
 \end{align}
 
 The Yangian Hopf algebra over the conformal algebra is generated by the above Lie algebra generators and an additional set of bi-local level-one generators of the form
 \begin{equation}\label{eq:lev1}
 \gen{ \widehat J}^A = f^{A}{}_{BC} \sum_{k=1}^n\sum_{j=1}^{k-1}\gen{J}_j^C\gen{J}_k^B +\sum_{j=1}^n v_j\gen{J}_j^A\,.
 \end{equation}
 Here $f^A{}_{BC}$ denotes the structure constants of the conformal algebra and the (a priori undetermined) variables $v_j$ parametrize an external automorphism of the Yangian. The level-one generators obey the commutation relations $\comm{\gen{J}^A}{\gen{\widehat J}^B}=f^{AB}{}_C \gen{\widehat J}^C$.
 
The structure of the conformal algebra ensures the full Yangian symmetry, as soon as invariance under a single level-one generator and the full level-zero algebra holds. A convenient choice for demonstrating this invariance is the level-one momentum generator $\gen{\widehat J}^A\sim \gen{\widehat P}$ given by
\begin{equation}\label{eq:Phat}
\gen{\widehat P}^{\mu}=-\ihalf\sum_{j<k=1}^n\big[(\gen{L}_j^{\mu\nu}+\eta^{\mu\nu}\gen{D}_j)\gen{P}_{k,\nu}-(j\leftrightarrow k)\big]
+\sum_{j=1}^n v_j \gen{P}_j^\mu.
\end{equation}
We may explicitly act with this generator onto the box integral \eqref{eq:boxint} to find
$
\gen{\widehat P}^{\mu} I_4=\sum_{j=1}^4 (v_j+j)\gen{P}_j^{\mu}I_4.
$
Hence, fixing the free parameters $v_j$ according to 
\begin{equation}\label{eq:choicev}
v_j^\text{box}:=-j,
\end{equation}
we see that the box integral is indeed Yangian-invariant. Parametrizing the box as
$I_4=\frac{1}{x_{13}^2x_{24}^2} \Phi(u,v)$ with 
the conformal cross ratios $u=\frac{x_{12}^2x_{34}^2}{x_{13}^2x_{24}^2}$ and $v=\frac{x_{14}^2x_{23}^2}{x_{13}^2x_{24}^2}$, this statement boils down to the following second order differential equation for the function~$\Phi$,
\begin{align}\label{eq:Phatinv}
0=\Phi
&+(3u-1)\frac{\partial \Phi}{\partial u}+3 v\frac{\partial \Phi}{\partial v}+(u-1)u \frac{\partial^2 \Phi}{\partial u^2}
\nonumber\\
&+ v^2 \frac{\partial^2 \Phi}{\partial v^2}+2 uv \frac{\partial^2 \Phi}{\partial u\partial v},
\end{align}
as well as the same equation with $u$ and $v$ interchanged.

Notably, the above box integral has a cyclic shift symmetry $x_k\to x_{k+1}$. For $v_j=0$ with $j=1,\dots,4$, this symmetry is violated by the level-one generators in \eqref{eq:lev1}. Crucially, the above non-trivial choice of parameters \eqref{eq:choicev} precisely restores this cyclic shift symmetry. In the case of the Yangian symmetry of tree-level amplitudes in $\mathcal{N}=4$ SYM theory (where $v_j=0$) this cyclicity is only possible due to the vanishing dual Coxeter number of the underlying Lie algebra $\alg{psu}(2,2|4)$. The above example thus shows that a vanishing dual Coxeter number is not necessary for the existence of cyclic Yangian invariants.

We will now show that Yangian differential equations, similar to \eqref{eq:Phatinv}, also hold for generic fishnet graphs.

\section{Scalar Fishnets and Monodromy}

Generic scalar fishnet graphs (cf.\ \figref{fig:fishnet}) are composed of the above box diagrams. 
In order to demonstrate their conformal Yangian symmetry, it is useful to formulate the above invariance in terms of the powerful RTT-formulation of the Yangian algebra. Here, the Yangian generators are packaged into a monodromy matrix
\begin{equation}
\mon(u)\simeq\idop +\frac{1}{u} \,\gen{J} +\frac{1}{u^2}\,\gen{\widehat J}+\dots,
\end{equation}
and the algebra relations are rephrased via the Yang--Baxter equation with Yang's R-matrix $\gen{R}(u)=\idop + u\, \permop$:
\begin{equation}\label{eq:RTT}
\rmat_{12}(u-v) \mon_1(u) \mon_2(v) = \mon_2(v) \mon_1(u) \rmat_{12}(u-v).
\end{equation}
We explicitly solve this RTT-relation by defining the monodromy as a product of conformal Lax operators \cite{Chicherin:2012yn}
\begin{equation}\label{eq:Lax}
\lax_{k,\alpha\beta}(u_k^+,u_k^-)
=u_k\idop_{k,\alpha \beta} +\sfrac{1}{2} \gen{S}_{\alpha\beta}^{ab} \,\gen{J}_{k,ab}^{\Delta_k},
\end{equation}
each of which obeys \eqref{eq:RTT} with $\mon_k\to \lax_k$. 
Here we package the inhomogeneities $u_k$ and the conformal dimensions $\Delta_k$ into the symmetric variables
$
u_{k}^+:= u_k +\frac{\Delta_k-4}{2}
$
and
$
u_k^-:= u_k-\frac{\Delta_k}{2}.
$
The $\gen{J}_{k,ab}$ denote the differential representation of the conformal algebra displayed in \eqref{eq:lev0gens}, and we have $\gen{S}^{ab}=\frac{i}{4}\comm{\Gamma^a}{\Gamma^b}|_\text{upper block}$, with $\Gamma^a$ representing six-dimensional gamma matrices for $\mathbb{R}^{2,4}$.
\begin{figure}[t]
\captionsetup[subfloat]{farskip=0pt,captionskip=0pt}
\subfloat[\label{fig:mongraph}]{
\includegraphicsbox[scale=.67]{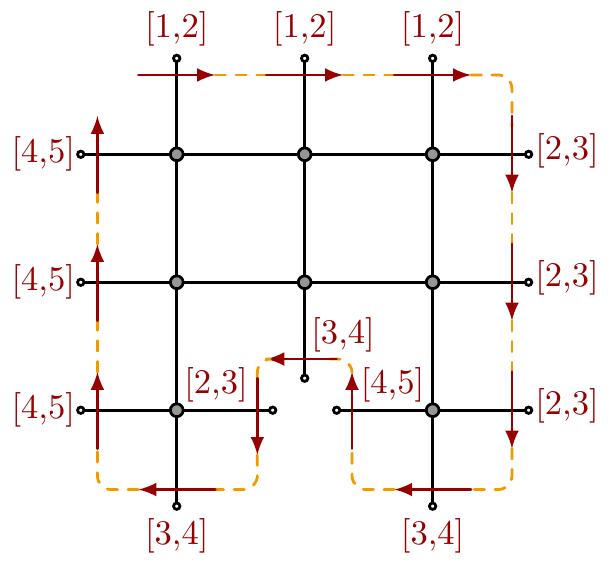}
}
\subfloat[\label{fig:mongraphrhs}]{
\includegraphicsbox[scale=.67]{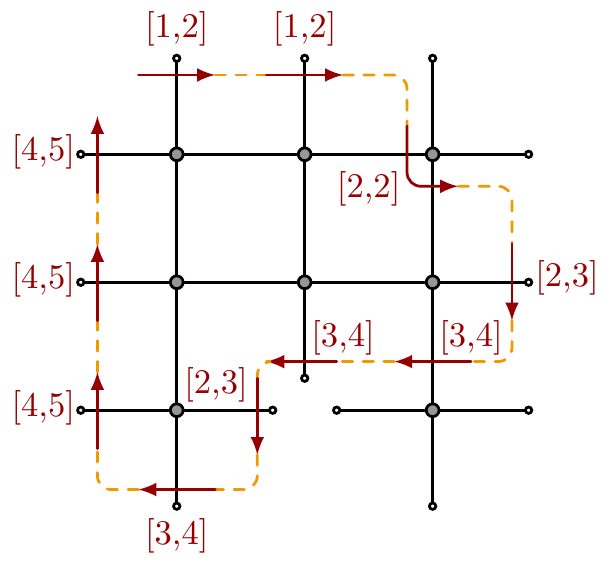}
}
\caption{
(a): Monodromy encircling a sample fishnet graph representing the left hand side of \protect\eqref{eq:moneigen}. 
(b): Intermediate step of the proof of Yangian symmetry.
}
\label{fig:mongr}
\end{figure}
The Yangian symmetry of the box integral $I_4$ and its $n$-point generalizations $I_n$ now translates into the eigenvalue equation~\cite{Chicherin:2013sqa}
\begin{equation}\label{eq:moneigen}
\mon(\vec{u}) \, I_n = \lambda(\vec{u})\, I_n \idop,
\end{equation}
where $\mon(\vec{u})$ denotes the inhomogeneous monodromy
\begin{equation}\label{eq:defmon}
\mon(\vec{u})=\lax_n(u_n^+,u_n^-)\lax_{n-1}(u_{n-1}^+,u_{n-1}^-)\dots \lax_1(u_1^+,u_1^-).
\end{equation}
The choice of parameters $u_k^\pm$ depends on the diagram under consideration. 
It will be convenient to introduce the notation $[\delta^+_k,\delta^-_k]:=(u+\delta^+_k,u+\delta^-_k)$ and $[\delta_k]:=u+\delta_k$.
By convention we choose the parameters on the boundary legs at the top to be~$[1,2]$. Then the parameters on the right, bottom or left boundary legs have to be $[2,3]$, $[3,4]$ or $[4,5]$, respectively (see \figref{fig:mongr} for an example).
The Lax operator defined in \eqref{eq:Lax} acts on an auxiliary and on a quantum space. While the product in \eqref{eq:defmon} is taken in the auxiliary space, each Lax operator acts on one external leg of the considered Feynman graph which represents the quantum space.
\begin{figure}
\captionsetup[subfloat]{farskip=0pt,captionskip=0pt}
\captionsetup[subfigure]{labelformat=empty}
\subfloat[\label{fig:elementruleslhs}]{
{\footnotesize (a) \hspace{2mm}} \includegraphicsbox[scale=.7]{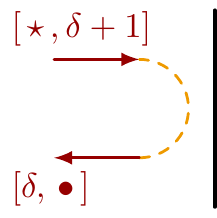}
$=$
\includegraphicsbox[scale=.7]{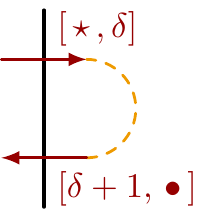}
}
\par
{\footnotesize (b)\hspace{2mm}} 
\subfloat[ \label{fig:elementrulesrhs}]{
\includegraphicsbox[scale=.8]{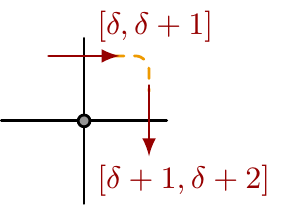}
$=$
\footnotesize$  [\delta+2]\times$
\includegraphicsbox[scale=.8]{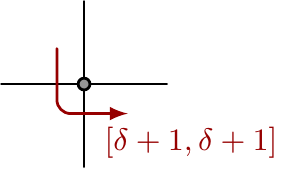}
}\par
{\footnotesize (c)\hspace{2mm}} 
\subfloat[ \label{fig:elementrule3}]{
\includegraphicsbox[scale=.8]{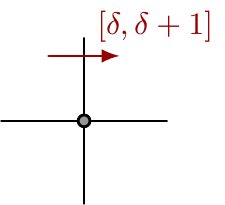}
$=$
\includegraphicsbox[scale=.8]{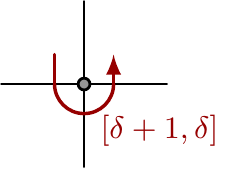}
}%
\caption{Rules employed to prove the Yangian invariance. 
(a): The intertwining relation \protect\eqref{eq:inter}.
(b) and (c): Pulling the monodromy contour through an integration vertex, cf.\ \protect\eqref{eq:rule2}.
}%
\end{figure}%

Proving the invariance statement \eqref{eq:moneigen} boils down to employing the lasso method \cite{Chicherin:2017cns}, i.e.\ to moving the monodromy through a given graph as displayed in \figref{fig:mongr}. The most important relation used in this process is the following intertwining relation for the Lax operator and the $x$-space propagator, cf.\ \figref{fig:elementruleslhs}:
\begin{equation}\label{eq:inter}
\frac{1}{x_{12}^{2}}\lax_2[\delta,\,\bullet\,]\lax_1[\,\star\,,\delta+1]=\lax_2[\delta+1,\,\bullet\,]\lax_1[\,\star\,,\delta]\frac{1}{x_{12}^{2}}.
\end{equation}
Moreover, we make use of the following relation, which allows us to move a product of Lax operators through an integration vertex, cf.\ \figref{fig:elementrulesrhs}:
\begin{align}\label{eq:rule2}
\int \dd^4x_0&\,\lax_2[\delta+1,\delta+2]
\lax_{1}[\delta,\delta+1]
\frac{1}{x_{01}^2 x_{02}^2 x_{03}^2 x_{04}^2}
\\\nonumber
&= 
[\delta+2]\int \dd^4x_0\,\frac{1}{x_{01}^2 x_{02}^2}\lax_0[\delta+1,\delta+1]\frac{1}{ x_{03}^2 x_{04}^2}.
\end{align} 
A third relation of this type is depicted in \figref{fig:elementrule3}.
 Finally, the Lax operator and its partially integrated version denoted by $\lax^T$ act on a constant function as
\begin{equation}\label{eq:rule3}
\lax_{\alpha\beta}[\delta,\delta+2] \cdot 1=\lax^T_{\alpha\beta}[\delta+2,\delta] \cdot 1 =[\delta+2] \delta_{\alpha\beta}.
\end{equation}
These rules are sufficient to move the monodromy contour in \figref{fig:mongraph} through the whole graph to end up with the eigenvalue on the right hand side of \eqref{eq:moneigen}. See \figref{fig:mongraphrhs} for an intermediate step.
The eigenvalue $\lambda(\vec{u})$ in \eqref{eq:moneigen} is composed of the factors picked up in this process via the relations \eqref{eq:rule2} and \eqref{eq:rule3}, cf.\ \cite{Chicherin:2017cns} for explicit expressions.


\section{Off- and On-shell Legs}

\begin{figure}
\includegraphics[scale=.8]{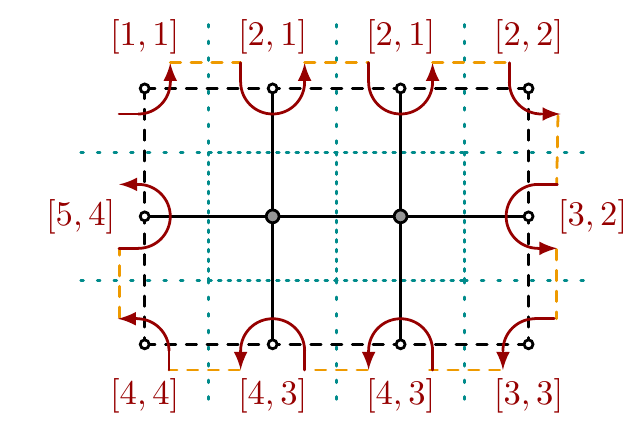}
\caption{
Double-box integral with massless external legs. Dashed black lines represent delta functions $\delta(x_{i,i+1}^2)$ forcing the momenta (dotted lines) on shell.}
\label{fig:doublebox}
\end{figure}

In the above construction, the external variables $x_i^\mu$ were in fact unconstrained.  To interpret the $x_i$  as region momenta for a scattering amplitude with massless on-shell legs, we require that 
$
p_k^2=(x_k-x_{k+1})^2=0.
$
Notably, the delta-function imposing this constraint obeys the same intertwining relation as the propagator in \eqref{eq:inter}:
\begin{equation}\label{eq:interdelta}
\delta(x_{12}^{2})\lax_2[\delta,\,\bullet\,]\lax_1[\,\star\,,\delta+1]
=
\lax_2[\delta+1,\,\bullet\,]\lax_1[\,\star\,,\delta]\delta(x_{12}^{2}).
\end{equation}
We may thus extend the above construction by introducing dashed lines alias delta functions into the graphs, see \figref{fig:doublebox} for the double-box example.
Due to \eqref{eq:inter} and \eqref{eq:interdelta}, propagators and delta functions are algebraically interchangeable. Hence, we can set external points (cf.\ \figref{fig:doublebox}) or internal propagators on shell via insertion of delta functions. 
Note, however, that the conformal symmetry of massless amplitudes typically shows an anomaly-like behavior for collinear configurations. This can be realized in subtle ways and may require additional contributions to the symmetry generators, cf.\ e.g.~\cite{Bargheer:2009qu,Sever:2009aa,Beisert:2010gn,CaronHuot:2011kk,Bargheer:2012cp}. The  investigation of this point is in progress.

The generalized boundary configurations as displayed in \figref{fig:doublebox} require to adapt the attribution of inhomogeneities. As can be seen for that example, the conformal dimension $\Delta_k=\delta_k^+-\delta_k^-+2$ entering the Lax operator  $\lax[\delta_k^+,\delta_k^-]$ still corresponds to the number of attached propagator lines, see \cite{Chicherin:2017cns} for details.

Notably, the central intertwining relation \eqref{eq:inter} generalizes to arbitrary powers $2\alpha$ of the propagator:
\begin{equation}\label{eq:intergen}
\frac{1}{x_{12}^{2\alpha}}\lax_2[\delta,\,\bullet\,]\lax_1[\,\star\,,\delta+\alpha]=\lax_2[\delta+\alpha,\,\bullet\,]\lax_1[\,\star\,,\delta]\frac{1}{x_{12}^{2\alpha}}.
\end{equation}
This allows us to construct Yangian-invariant deformations of the above correlators and amplitudes. These represent loop-level analogues of the tree-level amplitude deformations found in $\mathcal{N}=4$ SYM and ABJM theory~\cite{Ferro:2012xw,Bargheer:2014mxa}. 
Here the powers $\alpha_k$ of propagators entering a vertex have to obey the conformal constraint $\sum_{k}\alpha_k=4$. 

The theory defined by \eqref{eq:Lagr} is known to generate  double-trace interactions  \(\Tr(\phi_j\phi_j)\Tr(\phi_j^\dagger\phi_j^\dagger)\), \(\Tr(\phi_1\phi_2)\Tr(\phi_1^\dagger\phi_2^\dagger)\) and \(\Tr(\phi_1\phi_2^\dagger)\Tr(\phi_2\phi_1^\dagger)\) due to quantum corrections \cite{Sieg:2016vap}. It is important to note that these do not contribute  to the correlator \eqref{eq:tr-corr} at leading  order in \(N_c \) and hence to the considered planar observables.


\section{Including Fermions}

The procedure to obtain integrable quantum field theories as limits of $\gamma$-deformed $\mathcal{N}=4$ SYM theory suggests to extend our considerations to more general particle species. Adjusting the limit appropriately, one may for instance obtain an interaction Lagrangian including scalars and fermions \cite{Caetano:2016ydc}:
\begin{align}
\mathcal{L}_{\phi\psi}^\text{int}=N_\text{c} \Tr\big(&
\xi_1^2 \phi_3^\dagger \phi_1^\dagger\phi^3\phi^1
+
\xi_2^2 \phi_2^\dagger \phi_1^\dagger \phi^2\phi^1
\nonumber\\
&+
\sqrt{\xi_1 \xi_2}(\bar \psi_1 \phi^1 \bar \psi_4-\psi^1 \phi^\dagger_1 \psi^4)
\big).
\label{fermi-action}
\end{align}
Consider e.g.\ the following three-loop Feynman graph built from the above Yukawa vertices:
\begin{equation*}
\includegraphicsbox{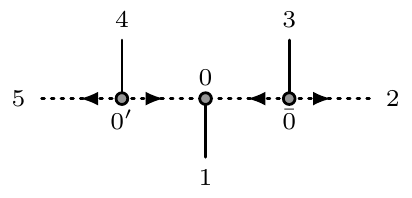}
\end{equation*}
Here dotted lines denote Fermion propagators. The corresponding integral reads
\begin{equation}
I_{\phi\psi}
=\int \dd^4 x_0 \dd^4 x_{\bar 0} \dd^4 x_{0'} \frac{
\rvec{P}{}_{\mu_2}^{\partial_{\tilde\mu_{\bar 0}}}
\vec{P}_{\partial_{\mu_0}}^{\tilde \mu_{\bar 0}} 
\rvec{P}{}_{\mu_0}^{\partial_{\tilde \mu_{0'}}} 
\rvec{P}{}_{\mu_5}^{\tilde\mu_{0'}}
}{x_{10}^2x_{3{\bar 0}}^2x_{4{0'}}^2},
\end{equation}
where the Fermion propagators are expressed using the notation
\begin{align}\label{eq:fermprop}
\rvec{P}{}_{A_1}^{A_2}&=\frac{\langle A_1| x_{12}|A_2]}{x_{12}^4},
&
\vec{P}_{A_1}^{A_2}&=\frac{[A_2| x_{12}|A_1\rangle}{x_{12}^4}.
\end{align}
We also employ dummy $\alg{su}(2)$ spinors with associated brackets $\langle\mu |$ and $[\tilde \mu |$, respectively, in order to avoid explicit spinor indices. Notably, also the propagators \eqref{eq:fermprop} obey intertwining relations including Lax operators $\lax^{\text{f}}$ and $\lax^{\overline{\text{f}}}$ in the $(\half,0)$ and $(0,\half)$ representations of the Lorentz group, respectively, for instance 
\begin{equation}
\lax_2^\text{f}(u+\sfrac{3}{2},\,\bullet\,)\lax_1(\,\star\,,u) \rvec{P}{}_{\mu_2}^{\partial_{\tilde \mu_1}}
=
\rvec{P}{}_{\mu_2}^{\partial_{\tilde \mu_1}} \lax_2(u,\,\bullet\,) \lax_1^{\overline{\text{f}}}(\,\star\,,u+\sfrac{3}{2}).
\end{equation}
For compactness, we refrain here from displaying the fermionic Lax operators; for details see \cite{Chicherin:xxxx}. Also in this case proving the Yangian invariance boils down to pulling the monodromy through propagators and vertices. The fermionic Lax operators are not proportional to the identity when acting on a constant  (cf.\ \eqref{eq:rule3} for the scalar case) but rather cancel via a relation \`a la $\lax^{\overline{\text{f}}}\lax^\text{f} P\sim P$ on the fermionic propagator $P$. 

If we consider the model \eqref{fermi-action} and an amplitude that describes the scattering of both types of fermions and only the boson \(\phi^1\), this scattering process is described by a single ``brick wall'' Feynman graph, whose bulk structure is given by a regular hexagonal fishnet lattice consisting of only Yukawa vertices, e.g.\
\begin{equation*}
\includegraphicsbox{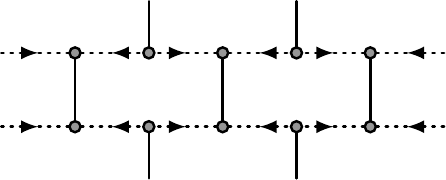}
\end{equation*}
 This type of integrable fishnet is new with respect to the examples given in~\cite{Zamolodchikov:1980mb}.  If we also include the other two types of bosons of \eqref{fermi-action}, each color-ordered amplitude is represented by a single graph with a mixture of rectangular
and hexagonal fishnet structures. 
These fishnet graphs also obey  the Yangian symmetry described above. 


\section{Three and Six Dimensions}

Instead of considering the parameter $\alpha$ in \eqref{eq:intergen} as a deformation in $4d$, we may directly associate it with the spacetime dimension $d$ via $\alpha=\frac{d-2}{2}$ and replace the four-dimensional Lax operators $\lax_k$ by an appropriate $d$-dimensional counterpart $\lax_k^d$, cf.\ \cite{Chicherin:xxxx}. 
Then the above off-shell construction generalizes to the cases of amplitudes in $d=3$ and $d=6$ spacetime dimensions built from scalar six- and three-point vertices, respectively.
The above action of the scalar Lax operator on the vacuum \eqref{eq:rule3} for these cases becomes
\begin{equation}\label{eq:rule3general}
\lax_{\alpha\beta}[\delta,\delta+\sfrac{d}{2}] \cdot 1=\lax^T_{\alpha\beta}[\delta+\sfrac{d}{2},\delta] \cdot 1 =[\delta+\sfrac{d}{2}] \delta_{\alpha\beta}.
\end{equation}
For  $d=3$ and $d=6$ the scalar graphs form triangular and hexagonal fishnets, respectively. These complete the set of regular tilings of the plane---all furnishing Yangian-invariant scalar Feynman diagrams, cf.\ \tabref{tab:overview}. Corresponding field theories were recently proposed in \cite{Caetano:2016ydc,Mamroud:2017uyz}. While the three-dimensional triangle graphs arise from scalar limits of planar, $\gamma$-deformed ABJM theory \cite{Caetano:2016ydc}, a six-dimensional ``mother'' theory is not known.

Due to the dimensionality of the propagators, in three and six dimensions we cannot use the naive trick to replace the propagator by a delta function $\delta(x_{ij}^2)$ in order to go on shell, cf.\ \tabref{tab:overview} and \eqref{eq:interdelta}. It is possible, however, to set up a momentum space Lax formalism in order to show the Yangian invariance of on-shell graphs~\cite{Chicherin:xxxx}.

\begin{table}[t]
\renewcommand{\arraystretch}{1.25}
\begin{tabular}{|l|c|c|c|}\hline
Dimension&$d=3$&$d=4$&$d=6$\\\hline
Propagator&$\abs{x_{ij}}^{-1}$&$\abs{x_{ij}}^{-2}$&$\abs{x_{ij}}^{-4}$\\\hline
\begin{minipage}[b]{1.08cm}\begin{flushleft}Scalar Fishnet \end{flushleft}\end{minipage}&
\includegraphicsbox[width=2.0cm]{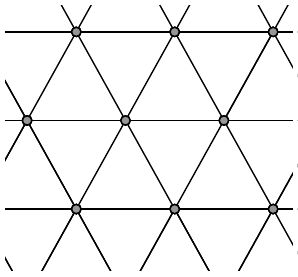}
&
\hspace{-.06cm}\includegraphicsbox[width=2.3cm]{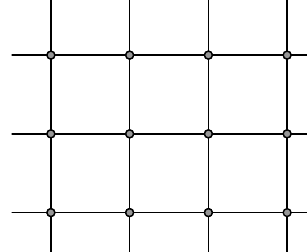}
&
\hspace{-.06cm}\includegraphicsbox[width=2.1cm]{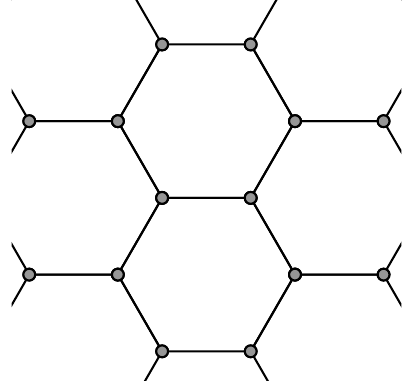}
\\\hline
\end{tabular}
\caption{Overview of scalar fishnet graphs.
}
\label{tab:overview}
\end{table}


\section{Conclusions and Outlook}

The Yangian algebra underlies the Bethe ansatz and the quantum inverse scattering method for the solution of rational integrable models. Its finding in the present context gives hope for the applicability of similar solution techniques to the largely unsolved class of fishnet integrals (cf.\ \cite{Basso:2017jwq} in this context). In four dimensions merely the box integral is solved~\cite{ussyukina1993approach} and already the double box is expected to yield complicated elliptic functions~\cite{CaronHuot:2012ab}. This renders new insights into the mathematical structure of fishnet integrals valuable.

The above classes of Feynman graphs define specific double scaling limits  of scattering amplitudes in  planar \(\gamma\)-deformed $\mathcal{N}=4$ SYM and  ABJM theory.  This suggests a set of non-trivial, integrable and non-supersymmetric CFTs in four dimensions, whose existence puts understanding the origins of integrability of the respective ``mother'' theories within reach.
This finding is also remarkable since the study of symmetry-invariant  subsectors has been crucial for developing the powerful integrability tools for the spectrum of AdS/CFT.
In particular, our results show that cyclic Yangian-invariant scattering amplitudes exist even if the dual Coxeter number of the underlying symmetry algebra does not vanish, i.e.\ for cases different from the full $\mathcal{N}=4$ SYM or ABJM theory. 

A further important goal
is to establish the Yangian symmetry for the most general double-scaled model of~\cite{Gurdogan:2015csr}, containing three couplings, three bosons and three fermions.


\acknowledgments{
We thank B.~Basso, J.~Caetano, L.~Dixon, J.~Henn, G.~Korchemsky and J.~Plefka for helpful discussions and comments.
This research was supported in part by the National Science Foundation under Grant No.\ NSF PHY-1125915.  The work
of V.K. and D.Zh. was supported by  the European Research Council (Programme  ERC-2012-
AdG 320769 AdS-CFT-solvable). V.K. thanks
Humboldt University (Berlin) for the hospitality and financial support in the framework of the
``Kosmos" programme.
}


\bibliography{YangianFishnetLetter}

\end{document}